# Dark Matter and Galactic Halos -A Quantum Approach

*A. D. Ernest*

School of Physical Sciences and Engineering

(University of New England, Armidale, Australia)

Traditional quantum theory can be used to construct hypothetical very large-scale gravitational stationary state structures from traditionally stable atoms and subatomic particles. These so called "gravitational macro-eigenstructures" have potential to explain the composition of extra-galactic dark matter and galactic halos. It is shown that the eigenstates within these structures can have radiative and stimulated lifetimes that are longer than the age of the universe, and also that they cannot be easily transformed or "destroyed" by many conventional galactic processes. Because of the unique nature of stationary states, it is shown that gravitational eigenstructures have the potential to remain largely undetected, provided certain conditions are met. Speculatively, it is suggested that they could provide a mechanism for the origin of high-energy cosmic rays, and also that, if these hypothetical structures have been present from an early time in the history of the universe, then they could have influenced the large-scale structure of the universe.

## 1. Introduction

The rotation velocity curves of stars in galaxies, the motions of pairs of galaxies and the behaviour of galaxies in clusters and super-clusters all indicate that the universe contains considerable quantities of non-luminous or dark matter. The debate over whether this matter is of a baryonic or non-baryonic form has continued for some time, the general consensus being that both types are present. Although to date no stable exotic non-baryonic particles have been detected, according to the standard model, element abundances determined from the nucleosynthesis era suggest that dark matter cannot be entirely baryonic (Silk 1995). By applying quantum theory to gravitational potentials on a large scale, this paper presents and investigates a further alternative: the feasible existence of large-scale, gravitationally bound eigenstates that are sufficiently "invisible" to be considered as possible candidates for dark matter. Such a structure will be referred to as a "gravitational macro-eigenstructure" (GME). The possibility will be investigated as to whether or not stable and almost totally dark eigenstructures could exist throughout the universe and contribute towards dark matter, and further, that GMEs might form the halos within which most or all galaxies are embedded.

Although the effects of quantum theory are generally ignored in all but the earliest times in the history of the universe, it is clear that the presence of universal eigenstates has the potential to influence such things as the development of the large-scale structure of the universe and the formation of galaxies. For example, the elemental ratios determined during the nucleosynthesis era assume and depend critically on, a uniform distribution of density of nucleons. Calculation of these ratios would clearly require the presence of matter in the form of macroscopic eigenstates to be taken into account.

Furthermore, the free particle or localised orbiting particle is a superposition of many eigenstates and reaction rates will be the average of that superposition. As will be demonstrated in this paper, the calculated reaction rates for interactions between specific individual eigenstates, or between individual eigenstates and free particles can therefore be quite different to those rates for interactions between localised free particles based on the average of the superposition.

It should be noted that this paper does not attempt to explain how or when these eigenstates might have developed. Nor does it try to trace galactic development based on the presence of gravitational eigenstates. Instead, it attempts to examine whether it is theoretically possible to form gravitational eigenstructures, and to ascertain the viability of gravitational eigenstates as dark matter candidates. Nevertheless, the speculative conjecture implicit in this paper is that the material universe may have existed partly, or even predominantly, in the form of (potentially macroscopic) stationary states (either gravitational or universal standing wave structures) from a very early time, and that these states may have coexisted with, and expanded concurrently with, the expansion of space, the formation of the galaxies occurring within eigenstructures to produce the visible galactic formations that are presently observed.

## 2. Formation of Macroscopic Eigenstructures

Quantum effects customarily manifest themselves only over atomic dimensions. There are however many examples of very large wave functions predicted directly from quantum theory. An *s*-wave photon emitted from a distant star may be light years across before it is destroyed or transformed via a transition within the retina of an observer. Individual visible-wavelength photons from a highly coherent laser can be many metres long. Particle wave functions exhibit similar expansions that can grow in size with time at a remarkably fast rate. The state vector probability function for an electron released from a cubicle box of side 1mm for example will fill an empty 1-litre container in about 1 1/2 seconds, while the corresponding wave function for an interaction initially constrained to nuclear dimensions, say 5 fm cubed, will reach galactic size within about 1000 years. For a proton this figure is about 500 000 years. (See appendix A.) There is no reason therefore to suppose that similarly large eigenfunctions could not exist.

The existence of a significant number of filled eigenstates on the macroscopic scale depends on:

1. the availability of suitably attractive potentials,
2. the existence of potentials that have a suitably large range, so that the eigenstates formed have appreciable range,
3. eigenvalues that are significantly negative, so that the particle or particles involved are sufficiently "bound", and
4. the susceptibility of the eigenstate (or lack of it) to radiative or induced transitions within the eigenstructure or to transitions induced by external influences (such as other particles traversing the eigenstate wave function volume).

It is customary to think of quantum mechanics as being associated with electrical or nuclear potential wells. To have a large effective radius, eigenstates associated with these potentials require high principal quantum numbers. Of course the difficulty is that, as the quantum numbers become large enough to produce macroscopic states, the energies associated with those states become so close to that of a free particle that the state could not remain stable. A hydrogen atom with n~3000 has an effective radius of about $10^{-3}$ mm and this corresponds to an energy of around one millionth of an eV.

For a gravitational potential, the simple two-particle Schrodinger equation is written as

$$-\frac{\hbar^2}{2\mu}\nabla^2\psi - \frac{GmM}{r}\psi = i\hbar\frac{\partial\psi}{\partial t}, \tag{1}$$

where M and m are the masses of the two particles, and μ is the reduced mass. Solutions to this equation yield the energies and wave functions of the gravitational eigenstates and are immediately obtained by comparison with those for the hydrogen atom (Schiff 1965) and give the energy eigenvalues $E_n$ as

$$E_n = \frac{-\mu G^2 m^2 M^2}{2\hbar^2 n^2} \tag{2}$$

and eigenfunctions $u_n(\mathbf{r},t)$ as

$$u_{n,l,m}(\mathbf{r},t) = R_{nl}(r) Y_{lm}(\theta,\phi), \tag{3}$$

where

$$\begin{aligned} R_{nl}(r) &= -\left\{\left(\frac{2\mu GmM}{n\hbar^2}\right)^3 \frac{(n-l-1)!}{2n[(n+l)!]^3}\right\}^{\frac{1}{2}} e^{-\frac{\mu GmMr}{n\hbar^2}} \left(\frac{2\mu GmMr}{n\hbar^2}\right)^l L_{n+l}^{2l+1}\left(\frac{2\mu GmMr}{n\hbar^2}\right) \\ &= -\left\{\left(\frac{2\mu GmM}{n\hbar^2}\right)^3 \frac{(n-l-1)!}{2n[(n+l)!]^3}\right\}^{\frac{1}{2}} e^{-\frac{\mu GmMr}{n\hbar^2}} \left(\frac{2\mu GmMr}{n\hbar^2}\right)^l \sum_{k=0}^{n-l-1} (-1)^{k+2l+1} \frac{[(n+l)!]^2 \left(\frac{\mu GmMr}{n\hbar^2}\right)^k}{(n-l-1-k)!(2l+1+k)!k!} \end{aligned} \tag{4}$$

Here $Y_{lm}(\theta,\phi)$ are the spherical harmonics, and $L_{n+l}^{2l+1}\left(\frac{2\mu GmMr}{n\hbar^2}\right)$ are the associated Laguerre polynomials.

Gravitational eigenstructures could conceivably consist of any stable particles such as protons, electrons, atoms or something more exotic. Using equations (3) and (4) and ignoring any attraction other than gravitational, two neutrons for example would have their first eigenstate spread over a distance greater than that of the galaxy and the energy eigenvalue of the system would be about $-10^{-69}$ eV, i.e., the eigenfunction becomes an essentially a free particle. On the other hand, two 1 kg "particles" would have a ground state energy at $-10^{65}$ eV but require a wave function of extent of only $10^{-58}$ m and exhibit the enormous density of $10^{58}$ kg $m^{-3}$. It is possible however to envisage several situations where, by choosing the appropriate mass, macroscopic wave functions can be formed with sensible energies. To do this requires large principal quantum numbers. For example, two $10^{-2}$ kg "particles" with $n = 10^{26}$ and $l = 10^{26} - 1$ would have a binding energy of about 6000 eV and an effective separation range of around 3 m. The traditional macroscopic situation of two orbiting masses however cannot be in any pure eigenstate as there are no eigenfunctions having this degree of localisation in the $\phi$ coordinate and the probability density is clearly not stationary. It would be virtually impossible to form macro-eigenstates of this magnitude here on Earth. Furthermore, the time required to observe a chance transition to this or similar states from two traditional orbiting masses would be prohibitively long because of the size of the relevant overlap integral.

In modelling dark matter two possible eigenstructure types are considered. The first is an attempt to apply the quantum eigenstate idea to crudely model the galactic halo by envisaging the visible and dark (eigenstructure) component behaving like a single central potential. Each particle (a proton, electron, atom or something more exotic) exists in an eigenstate around this central potential with its wave function at a large radius (like an orbiting particle in the galactic halo but in the form of an eigenstate rather than a mixed state that forms a traditional localised orbiting particle).

In the second, an essentially totally dark structure (or "dark" galaxy) is created by having the very large number of small mass particles (again protons, electrons, atoms and/or more exotic matter) all bound together as a single quantum eigenstructure without any discrete visible massive central core.

A. Gravitational Macro-Eigenstructures and the Galactic Halo

A very gross two-particle approximation of the galactic halo may be made by treating each particle within the halo as acting independently and subject to a single central gravitational potential produced by a mass of $10^{42} – 10^{43}$ kg. As indicated above, this halo particle could be anything intrinsically stable (proton, electron, atom etc.). Eigenfunctions for a particle of mass $\sim10^{-27} – 10^{-31}$ kg influenced only by gravity are examined. There are clearly some extreme simplifications involved here. In reality the bulk of the matter consists itself of eigenstates and resides within in the halo. Its mass distribution has however been treated as spherical so that it can still behave as a single central potential from the point of view of the eigenstate particle under consideration.

The dynamic and often violent natures of many processes that occur within the galaxy and its neighbourhood have been assumed to have a negligible effect. The justification for adopting this assumption is the notion that an approximation equivalent to the adiabatic one used with electronic and vibrational motions in atomic physics, should be applicable here, provided that the particle eigenfunction has a relatively small mass and completes its oscillations in a sufficiently short time. This is the case, since the quantity $\hbar/E$ in the time dependent part of the eigenfunction ($e^{-iEt/\hbar}$) is short compared to most large-scale galactic processes and clearly the mass is small. The eigenstate wavefunction should therefore adjust to the (relatively) "slow" changes in the mass distribution of the galactic core in the same way that electron eigenstate wavefunctions adjust to vibrations of the ion cores of atoms. This should also mean that despite temporal changes in galactic mass distribution, the eigenstates should remain pure rather than evolving into mixed states.

Overall electrical neutrality requires that there be equal numbers of electrons and protons in the halo. No attempt has been made to include any residual electrical effects, such as those due to near neighbours* or the electrical effects due to the differences in the decay rates of each of these particles. It is nevertheless useful as a first step to consider the simple model and examine the stability and longevity of the eigenstates with respect to aspects such as spontaneous and stimulated decay, interaction with

*The electrical potential energy of a single proton of charge $q$ embedded in a uniform array of galactic positive and negative charges $q_j$ is given by $\frac{1}{4\pi\varepsilon_0}\sum_{\text{all charges } q_j}\frac{(\pm)qq_j}{r_j}=\frac{\alpha q^2}{4\pi\varepsilon_0 r_j}$, where $\alpha$ is the Madelung constant (Kittel 1971). It is shown later that the density of eigenstate particles is of the order of $10^6$ m$^{-3}$ or less. In anticipation of this result the particle separation is taken to be $10^{-2}$ m. This gives roughly $\sim 10^{-26}$ J for the electrostatic potential energy while the gravitational potential energy of the proton would be $\frac{GmM}{r}=\frac{6.7\times10^{-11}\times1.7\times10^{-27}\times10^{42}}{10^{21}}\sim 10^{-16}$ J and is the dominant term.

radiation and visible matter, and the ability of these states to remain undetected. For such a system, the state energy, the extent of the wave function, the mass stored in the form of eigenstates, the radiative decay rate, the susceptibility to induced transitions, the interaction with traditional matter and other factors are addressed below. Of course the eigenstates of the simple model will not be true eigenstates of a real galaxy (but a superposition of them). There would however be an equivalent set of eigenfunctions for the real galaxy with properties presumably not too different from those examined in the model.

*(i) Eigenstate energy and extent of the wave function*

For the masses considered, the eigenfunctions that still have reasonable binding energies encompass distances of the order of $10^{20}$ - $10^{21}$ m and have very large principal quantum numbers, ($n$~$10^{30}$). For the present discussion the significant part of the wave function (3) above is the dependence on $r$ of the form of the radial component $R_{nl}$ of the wave function $u_n(\mathbf{r},t)$ and the $r$ dependence of its associated probability density $(R_{nl}r)^2$. The mathematical complexity of the Laguerre polynomials for $n \sim 10^{30}$ makes it virtually impossible to study their properties directly. It is possible however to make some empirical generalisations about the radial eigenstate wave functions. In particular, their radial extent for large $n$ and $l = 0$ is given in appendix B. Using equation (B2) that

$r_{\text{finalmax}} \approx 2n^2 \dfrac{\hbar^2}{\mu G m_p M}$, where $M$ = galactic mass = $10^{43}$ kg, $r_{\text{finalmax}} \approx$ radial extent of wave function = 5 x $10^{21}$ m and $m_p$ = proton mass = 1.67 x $10^{-27}$ kg, we get $n \approx$ 8 x $10^{33}$ and this gives, using equation (2), estimates of the binding energy $E$ of a proton in a gravitational eigenstate of about –250 eV. States with smaller *n* values and wave functions closer to the galactic centre are of course much more strongly bound. For an electron *n* and *E* are smaller by about a factor of 2000.

*(ii) Mass capacity of a halo eigenstate*

Given that the value of *n* might be as high as $10^{34}$, how much matter could be stored in such eigenstructures? There are *n* values of *l* for each *n*, and (2*l* + 1) values of *m* for each *l*. For fermions then, there are a total of $\frac{1}{3}n(n+1)(2n+1)$ possible states for *n* = 1 to *n*. For $n=10^{34}$ this corresponds to about $10^{101}$ states or $10^{74}$ kg for eigenstructures made of baryonic type particles (or somewhat less if the eigenstates are neutrinos). Even if only the high *n* states (say $n = 10^{33}$ to $n = 10^{34}$) and only the very high *l* values of those states (say $l = n$ to say $l = n - 10^{6}$) are considered, there are clearly plenty of states available, since the mass of the galaxy is estimated at about $10^{42}$ or $10^{43}$ kg including the dark matter.

*(iii) Radiative decay*

Radiative decay of gravitational macro-eigenstates could take place via the emission of gravitational and/or electromagnetic wave radiation if the eigenstate particles are charged. There is no full quantum theory of gravity as yet. Nevertheless it is easy to show classically that radiative decay through gravity wave radiation is almost certainly insignificant compared to that produced through electromagnetic radiation. From classical electromagnetic theory, equations for the electric field strength $\mathbf{E}_{\mathbf{rad}}$ of the electromagnetic radiation field, and the total electromagnetic radiated power *P*, found from integrating the Poynting vector **S** over a spherical surface of radius *r*, are given respectively by, in the low velocity case (Panofsky and Phillips 1969)

$$\mathbf{E}_{\mathbf{rad}} = \frac{e}{4\pi\varepsilon_0 r^3 c^2}\mathbf{r}\times(\mathbf{r}\times\dot{\mathbf{u}}) \tag{5}$$

and

$$P = \frac{e^2\dot{u}^2}{6\pi\varepsilon_0 c^3} \tag{6}$$

Here, $\dot{\mathbf{u}}$ is the acceleration of the particle, r is the radius of the spherical surface and the other terms have their usual meanings. In an analogous way equivalent expressions for the gravitational wave radiation field $\Gamma_{\mathbf{rad}}$ and the gravitational radiated power *P*, found from integrating the gravitational equivalent of the Poynting vector over a spherical surface radius *r*, may be written as

$$\Gamma_{\mathbf{rad}} = \frac{Gm}{r^3 c^2}\mathbf{r}\times(\mathbf{r}\times\dot{\mathbf{u}}) \tag{7}$$

$$P = \frac{2Gm^2\dot{u}^2}{3c^3} \tag{8}$$

Comparing say, for a proton, the value of the term $\frac{e^2}{6\pi\varepsilon_0}$ of equation (6) with $\frac{2Gm^2}{3}$ in equation (8) respectively, shows that the rate of energy loss via electromagnetic radiation is about $10^{37}$ times larger than that via gravitational radiation.

The forgoing calculation suggests that it is only necessary here to consider electromagnetic decay rate. In this case analogous equations to those in atomic physics may be used provided that similar assumptions are made about the distribution or the radiation field. The transition rate *TR* for a given transition $n_i \rightarrow n_f$ is given by (Corney 1986)

$$TR = \frac{\omega^3 {p_{if}}^2}{3\varepsilon_0 \pi \hbar c^3}, \qquad (9)$$

where $\omega = \frac{\mu G^2 {m_p}^2 M^2}{2\hbar^2}\left(\frac{1}{{n_f}^2} - \frac{1}{{n_i}^2}\right)$ is the angular frequency corresponding to the transition *i* to *f*, $\mu$ is the reduced mass, $p_{if}$ is the corresponding dipole matrix element for the transition and the other symbols have their normal meanings.

There are many more states available than is required to provide the amount of dark matter estimated to be present in the galactic halo. It is not possible to write down a completely general expression for the dipole matrix elements for all the different possible transitions. In traditional atomic physics the high *n*, high *l* states are the ones with the longest lifetimes so if stable, long lived gravitational eigenstates exist, they are most likely to be ones with similarly high *n* and *l* values. Figure 1 shows the high *n*, high *l* values being considered here. Rather than taking space here, an outline of the derivation of some of the relevant $p_{if}$ values and their incorporation into the transition rates is given in appendix C. Nevertheless the crux of the hypothesis that gravitational eigenstates might be candidates for dark matter rests on the unusual properties of this specialised subgroup of states shown in figure 1, and therefore the relevant and important points from this appendix are summarised here:

(1) As radiative decay proceeds, there is an average net migration of states towards the left hand diagonals, i.e., to those exhibiting the highest angular momentum and longest lifetimes.

(2) The lifetimes of these high $l$ states are extremely long, in general many times the age of the universe. States on at least the first billion diagonals of figure 1 have these extreme lifetimes.

(3) The critical factor in determining the matrix element $p_{if}$ is its radial component. This radial component decreases dramatically as the difference in principal quantum number $\Delta n = n_f - n_i$ becomes large. This decrease far exceeds any increase in transition rate due to an increase in $\omega$. (See appendix C for an explanation of this effect and a table of typical values).

The transition rate for a particular eigenstate is then determined by substitution of $\omega$ and $p_{if}$ into equation (9) with the $p_{if}$ values determined by the relevant angular and radial integrals of appendix C. As an example, transitions of the type A to A′ in figure 1 have $n_f = n_i - 1$ and $j = 1$. Taking a central galactic mass of $10^{42}$ kg, and a value of $n_i$ of $4\times10^{30}$, an electron wave function would have a radius of about $10^{21}$ m, a radial spread of around $5\times10^{6}$ m and a transition rate of $4\times10^{-25}$ $s^{-1}$ (or radiative lifetime of $2.5\times10^{24}$ s). This rate is of course extremely slow and the state is effectively frozen in time. To decay to $n = 1\times10^{30}$ requires $3\times10^{30}$ jumps or a total time of $6\times10^{54}$ s or $2\times10^{47}$ years. This is far longer than the lifetime of the universe, which is around $5\times10^{17}$ s. For protons, the principal quantum number that gives a radius of around $10^{21}$ m is $n_i = 8\times10^{33}$ For this value of $n_i$, the wave function spread is about $10^5$ m and the single transition lifetime is about $4\times10^{24}$ s. Again the lifetimes are far older than the age of the universe. In fact, because of point (3) above, all states on at least the first billion or so diagonals of figure 1 are effectively frozen in time and high angular momentum halo eigenstates will clearly not decay in any significant time scale.

*(iv) Stimulated emission and absorption of radiation*

It has been demonstrated in the previous section that some of the gravitational eigenstates of the type envisaged in this paper have extremely long radiative decay times. There is also the matter of the stability of the eigenstate with respect to absorption of radiation. The cosmos is filled with radiation, particularly within, and in the vicinity of, a galaxy. The potential exists for eigenstates to be shifted up or down via stimulated absorption or emission of radiation. An eigenstate could be successively promoted upwards by radiation until it is ultimately gravitationally freed from the galaxy, in a process analogous to the electrical ionisation of an atom. Cosmic background radiation (CBR) might conceivably do this.

The rate of absorption of radiation (or stimulated emission) from a state 1 to 2 is given by $B_{12}\rho(\omega)$ where $B_{12}$ is the Einstein *B* coefficient and $\rho(\omega)$ is the spectral energy density of the radiation with respect to $\omega$ (Corney 1986). The spectral energy density per unit angular frequency of CBR has a maximum value of approximately 3 x $10^{-26}$ J s $m^{-3}$ at an angular frequency of 2 x $10^{12}$ $s^{-1}$ for a 3K background temperature. Now if the background radiation is in equilibrium with the macroscopic eigenstructure then $B_{12}\rho(\omega) = \frac{A_{12}}{\exp\left(\hbar\omega/_{kT}\right)-1}$ where $A_{12}$ is the spontaneous emission rate given by equation (9) and $\exp\left(\hbar\omega/_{kT}\right)-1$ has a value of about 20 at $\omega =$ 2 x $10^{12}$ $s^{-1}$. For this angular frequency the corresponding change in principal quantum number is given from equation (2) as $\Delta n = \frac{\hbar^3 n^3 \omega}{\mu G^2 m^2 M^2}$, (provided $\Delta n << n$). For a particle whose mass is similar to that of a proton, this gives $\Delta n \sim 10^{28}$ while for an electron $\Delta n \sim 10^{26}$. Again the significant quantity in the transition rate $A_{12}$ is the factor $\omega^3 {p_{if}}^2$. The angular frequency of radiation has increased by a factor of around $10^{30}$ compared to that for a transition with $\Delta n = 1$, but from table 1 $p_{if}$ has decreased by a far greater amount, so that the product $\omega^3 {p_{if}}^2$ is still very much reduced. It is clear that to promote a state from $(n_i, l_i) = (n_i, n_i - 1)$ to $(n_f, l_f) = (n_i + 10^{26}, n_i)$ will be impossible in the lifetime of the universe. For transitions originating from states with $l \neq n-1$ (*i.e.* $j \neq 1$), similar orders of magnitude for $p_{if}$ should apply, provided that $j << n$. The results will certainly apply for initial states on any of at least the first billion diagonals of figure 1. Inelastic scattering of CBR photons with eigenstates is therefore minimal.

The derivation of the overlap integral is contingent on the radiation field being uniform over the entire wave function. In scenarios involving traditional atomic transitions the electromagnetic field invariably behaves in this way because the electronic wave functions are generally embedded within the field. Indeed the term $\exp(\mathrm{i}\mathbf{k} \cdot \mathbf{r})$ in the transition probability formula $R_{ik} = \frac{\pi e^2}{2\hbar^2} \left| \langle k | \mathbf{r} \cdot \mathbf{E}_\omega \exp(\mathrm{i}\mathbf{k} \cdot \mathbf{r}) | i \rangle \right|^2$ (Corney 1986) is expanded as a Taylor series on the basis that $|\mathbf{r}|$ is small over any region where the wave functions are significant so that that $\mathbf{k} \cdot \mathbf{r}$ is much less than unity. With macroscopic eigenfunctions however it is quite feasible that a relatively localised photon wave function may only occupy a small fraction of the region of space occupied by the eigenstate wave functions and the assumption of putting $\exp(\mathrm{i}\mathbf{k} \cdot \mathbf{r}) = 1$ is not justified. The value of the overlap integral may therefore be considerably different to that calculated with the photon field spread uniformly over the eigenstates. Statistically however this would not be a problem for CBR because of its presumed uniformity throughout space. When considering radiation from the galaxy however, different spatial regions within the particle wave function are exposed to different intensities and spectral qualities of the electromagnetic photon field. The electromagnetic radiation field intensity near a stellar surface may be very great, but not sufficient to induce transitions simply because it interacts with an insignificantly small part of the particle eigenfunction. Nevertheless what is of importance here is to determine (albeit very approximately) whether there is any chance that galactic radiation could influence macroscopic eigenstructures. A calculation based on the average radiation that would pass through an eigenstate due to galactic emission should therefore be considered. The galactic radiation field is treated as uniform across the eigenstate and the assumption that $\exp(\mathrm{i}\mathbf{k} \cdot \mathbf{r}) = 1$ is maintained, which although clearly not correct, should still serve to give an estimate of the effects of such radiation.

The average energy output of the Sun is 3.6 x $10^{22}$ W and the luminosity of the galaxy is rated at $10^{10}$ times the solar output. The galaxy is approximated as a radiating black body of total energy output of 3.6 x $10^{32}$ W with temperature of 8000 K (a typical stellar surface). Using the Stefan-Boltzmann relationship that the radiancy, $R = \sigma T^4$ where $\sigma$ is the Stefan-Boltzmann constant and *T* the temperature, this power corresponds to a black body sphere of radius 8 x $10^{12}$ m. It is assumed that this sphere of radiation is small compared to the radial position of the eigenstate ($r$ = $10^{20}$ m) and so behaves as an equivalent point source. The integrated energy density $\int_0^\infty \rho(\omega)d\omega$ at radius $r$ = $10^{20}$ m is then $\frac{10^{32}}{4\pi r^2 c} \sim 10^{-17}$ J m$^{-3}$. Since the spectral quality is assumed to be that of a black body, and further that this spectral quality does not alter significantly on its outward journey, $\rho(\omega)$ across the *entire* eigenfunction is approximated as $\frac{A\hbar\omega^3}{\pi^2 c^3\left(\exp\left(\hbar\omega/kT\right)-1\right)}$ where $A$ could be considered as a fractional energy density dilution factor for the black body radiation as it radiates out into the halo, to be determined from the total integrated energy density. Performing the integration and taking $r$ as $10^{19}$ m (having $r$ as a minimum position for the start of the wave function ensures a maximum possible value of $\rho(\omega)$), gives $A$ as approximately 8.5 x $10^{-17}$. Reference (Corney 1986) gives a general formula for the rate of stimulated emission and absorption of unpolarised, isotropic radiation as $R_{ik} = \frac{\pi e^2}{3\varepsilon_0\hbar^2}\left(\left|\langle k|x|i\rangle\right|^2 + \left|\langle k|y|i\rangle\right|^2 + \left|\langle k|z|i\rangle\right|^2\right)\rho(\omega)$. The change in quantum number $\Delta n$, between the states *k* and *i* depends on the frequency $\omega$. Wein's Displacement Law gives the maximum value of $\rho(\omega)$= 3.4 x $10^{-32}$ J m$^{-3}$ at $\omega$ = 5 x $10^{15}$ $s^{-1}$ for $T$ = 8000 K. $\omega$ = 5 x $10^{15}$ $s^{-1}$ corresponds to a of $\Delta n \sim 10^{31} - 10^{32}$ for protons and $\Delta n \sim 10^{29} - 10^{30}$ for electrons. The quantities $\mathrm{e}\left|\langle k|x|i\rangle\right|^2$, $\mathrm{e}\left|\langle k|y|i\rangle\right|^2$ and $\mathrm{e}\left|\langle k|z|i\rangle\right|^2$ relate directly to the radial part of the overlap integral $\int_0^\infty eR^*_{nf,lf}(r)r^3R_{ni,li}(r)r^2\,dr$ and have essentially been calculated in appendix C for transitions of the type $n_i, l_i = n_i - j$ to $n_f = n_i - j, l_f = n_f - 1$ (or vice versa). The final row of Table 1 demonstrates the overwhelming influence of $\int_0^\infty eR^*_{nf,lf}(r)r^3R_{ni,li}(r)r^2\,dr$ on the size of the overlap integral and that such stimulated emission and absorption rates are negligible. Also note that, even though they have not been explicitly calculated, rates for transitions originating from other diagonals are also likely to be extremely small, provided that $j << n$ for the initial state. Even near a supernova explosion, where $\rho(\omega)$ has the intensity many orders of magnitude larger than those just discussed, the photon density is large only over a relatively very small volume and it can be estimated that the ability to induce transitions within the eigenstates remains insignificant.

*(v) Interaction with visible matter*

Examining whether or not bulk visible matter such as a gas/dust cloud or stars could influence the density of eigenstructure particles from a quantum mechanical viewpoint is clearly a complex process. Instead a more simplified estimate is undertaken here based on the correspondence principle. The assumption is made that as stars sweep out a volume of space eigenstate particles make transitions to, and become part of, the stellar material. Furthermore it is assumed that the number of eigenstates removed from the eigenstructure is directly proportional to the volume swept out by the stars. The volume swept out per year by a typical star like the sun is the cross sectional area of the star multiplied by the distance travelled in one year. Multiplying this volume by say $10^{11}$ for the number of stars in the galaxy gives a total volume swept out by all stars in the galaxy as 6 x $10^{42}$ $m^3$ per year. Taking the extent of the eigenstructure halo as $r = 10^{21}$ m, the density of eigenstate material within the galactic halo would be about 2 x $10^{-22}$ kg $m^{-3}$ or $10^5$ proton mass particles per cubic metre. Hence approximately $10^{47}$ particles would be removed from the eigenstate per year or $10^{57}$ particles in the lifetime of the universe. Since there are $10^{70}$ particles making up the eigenstructure this is an insignificant amount. This is probably an overestimate since stars predominantly orbit the near the galactic plane, and as such those eigenstates with *m* values that have high eigenfunction densities within the galactic plane would be preferentially removed. A similar argument applies for gas and dust clouds, that is that the eigenstates within the galactic plane would be removed leaving an extended halo above and below the plane of rotation of the galaxy.

*(vi) Recombination of proton/electron eigenstates*

One possibility offered here regarding the composition of the dark halo has been that it consists of equal numbers of protons and electrons (rather than hydrogen atoms) whose electrical interactions have been assumed, in general, to cancel, leaving gravitational attraction as the predominant attractive potential. To this end it requires that the particles do not recombine on time scales that are short compared with the lifetime of the galaxy. The rate of simple two body radiative recombination in a neutral plasma in thermal equilibrium is given by $\frac{\partial N}{\partial t} = -\alpha N^2$ which has solution $\frac{1}{N} = \frac{1}{N_0} + \alpha t$ where $N$ is the electron\ion density at time $t$, $N_0$ is the initial electron\ion density and $\alpha$ is the recombination coefficient. $\alpha$ is related to the velocity dependent, recombination cross section $\sigma(v)$, via an integral over the velocity distribution function. $\sigma(v)$ is related to the rate of change in number density $\frac{dN_1}{dt}$ of beam particles of velocity $v$ relative to the target particles of density $N_2$, is given by

$$\frac{dN_1}{dt} = N_1 N_2 \sigma v \text{ or } \frac{dN_1}{dt} = kN_1 \text{ where } k = N_2 \sigma v \tag{10}$$

It was explained earlier that for a given radius, the principal quantum number $n$ and state energy $E$ for electrons is about 2000 times smaller than for protons. Now it is known from classical considerations the (negative) potential energy of the bound particle is twice as large as its (positive) kinetic energy. This means that a proton and an electron in a typical high $l$ state whose eigenfunctions happen to occupy the same or almost the same region of space (with $r$ say about $10^{20}$ to $10^{21}$ m) will exhibit a difference in their kinetic energy of about 250 eV. The eigenstates are clearly not in thermal equilibrium with each other and it is therefore more appropriate to use $\frac{dN_1}{dt} = N_1 N_2 \sigma v$. If $N_1 = N_2 = 10^5$ m$^{-3}$, the relative velocity $v$ (corresponding to 250 eV) is taken as $10^5$ m s$^{-1}$ and $\sigma(v)$ is estimated by extrapolation of the data of (Papoular 1965) as approximately 2 x $10^{-27}$ m$^2$ then the change in number density is given by $\frac{dN}{dt} = 2 \times 10^{-17} N$, indicating that the recombination time is less than, although of the order of, the age of the universe. There is of course considerable uncertainty in this result, because of the estimate of $\sigma$ and the number densities. It is therefore difficult to say whether this result suggests that the eigenstates would be in the form of protons and electrons or as neutral atoms. It does however point to the possibility that, if dark matter galactic halos are in the form of these types of eigenstates, then their detection via recombination radiation may be possible. Indeed there is already some evidence of the emission of radiation from some galactic halos. (See for example page 232 of (Silk 1995).)

*(vii) Photo-ionisation and photo-excitation of neutral hydrogen*

Photons originating external to or from within our own galaxy have potential to be elastically or inelastically scattered from the eigenstate particles. If eigenstates were in the form of atomic hydrogen (at density $10^5$ m$^{-3}$), then photo-excitation and photo-ionisation are possible. The flux density of photons originating from within our own galaxy is by far the more intense source. If it is imagined that the photons of density $N_1$ from the galaxy continually pass through a given volume containing eigenstate atoms at density $N_2$ and continually ionise them then equation (26) gives the rate of re-ionisation of any hydrogen atoms formed through recombination as $\frac{dN_2}{dt} = N_1 N_2 \sigma c$. Taking the energy density of photons suitable for ionisation (frequencies > 15 eV) as $\rho(\omega) \sim 10^{-32}$ J m$^{-3}$ gives a photon number density of $N_1 \sim 10^{-14}$ m$^{-3}$ and hence the re-ionisation rate $\frac{dN_2}{dt}$ as $10^{-22}$ m$^{-3}$ s$^{-1}$. This small rate compared to the suggested rate for recombination given in the previous section suggests that 're-ionisation' of any of recombined eigenstate particles will not happen in the present lifetime of the universe.

Equation (10) can also be rewritten by noting that the photon number density can be related to the photon energy density $\rho(\omega)$ and photon intensity *I* by $N_1 = \frac{\rho(\omega)}{\hbar\omega} = \frac{I}{c\hbar\omega}$ to give $\frac{dI}{dx} = -N\sigma I$ with solution $I = I_0 \exp(-\sigma N x)$. If the atomic density *N* is taken again as $10^5$ m$^{-3}$ and $\sigma$ for photo-ionisation is taken as ~ $10^{-21}$ m$^{-2}$ (or ~ $10^{-22}$ m$^{-2}$ for photo-excitation) then $I = I_0 \exp(-10^{-16} x)$. For photons traversing an entire eigenstructure of distance $x = 10^{21}$ m say, this equation suggests there would clearly be significant attenuation of any light that was capable of causing photo-ionisation or photo-excitation. The fact that there appears to be no evidence of the attenuation of radiation from galaxies at the excitation frequencies of hydrogen or beyond the ionisation threshold frequencies suggests that eigenstructures would consist predominantly of material at least as elementary as protons and electrons rather than in the form of atoms. This in turn puts an upper limit on the density of eigenstate material in the halo and on the ability to detect it by recombination radiation.

*(viii) Elastic (Thomson) scattering of photons from eigenstate material*

Photons are easily scattered off free electrons. If the eigenstructure material is of the form of ionised hydrogen, then one might at first expect that substantial scattering off free electrons could occur and that eigenstates would be detectable through this scattering. This is however not likely, for the following reason. In traditional atomic structures, elastic scattering does not occur from eigenstate electrons bound in atoms because the atom must recoil as a whole, resulting in a negligible elastic scattering cross section. Similarly, electrons in gravitational eigenstates are bound to the galaxy as a whole and the galaxy must recoil as a whole. The only possible motion of the electron is via a gravitational eigenstate transfer through inelastic scattering, which has already been discussed in (*iv*) above. This observation illustrates one of a number fundamentally important differences in behaviour that can exist when comparing the interaction rate properties of localised particles consisting of a superposition of states with those of eigenstates.

*(ix) Inelastic (Compton) scattering of photons from eigenstate material*

Although it was seen that the photon-stimulated transition rates between eigenstates are negligible, Compton scattering could cause the eigenstate to become mixed or alternatively "gravitationally ionise" the particle from the galaxy altogether. If the mean density of x-rays is $N_x$ and that for eigenstate particles is $N$ then the reaction rate is again given by a modification of equation 10 as $\frac{dN_x}{dt} = \frac{dN}{dt} = N_x N\sigma c$, where $c$ is the speed of light and $\sigma$ the Compton cross section. If it is assumed that each time a reaction occurs it results in a conversion of an eigenstate to a mixed state (or alternatively results in a gravitational ionisation event), then an effective "mixing time" (or ionisation time) $\tau_m$ can be defined as $\tau_m = \frac{N}{dN/dt} = \frac{1}{N_x \sigma c}$. Taking the maximum value of $\sigma$ as 6.7 x $10^{-29}$ $m^2$, $N$ as $10^5$ $m^{-3}$ and $N_x$, derived from the local solar x-ray flux, as $10^2$ $m^{-3}$ (the average galactic x-ray photon density being orders of magnitude lower than this), gives the mixing time as at least the age of the universe. The conclusion is that Compton scattering cannot significantly shift the eigenstates into mixed states or gravitationally ionise them. In any case an argument similar to the recoil one in the previous section should still apply implying that the cross section for Compton scattering should be based on the mass of the whole galaxy and hence be much lower than that given above.

B. Eigenstructures with no visible component

It is theoretically possible to have a gravitational eigenstructure that has no visible component. There are several examples of observations that indicate the presence of objects that exhibit gravitational forces but are not seen, for example the so-called "great attractor" and cases of gravitational lensing where the intervening galaxy presumably responsible for the lensing process cannot be detected (Silk 1995). A halo of eigenstates surrounding a non-visible massive centre, e.g., a black hole, would be a possibility. A structure could however be formed entirely of eigenstates. For a small number of particles, the states are widely spread and have too little binding energy too be stable. As more particles are added however, the binding energy of the states increases and the structure becomes more compact. It is first assumed that the eigenstructure consists of a total mass $M$ (= $\sum m_i$ ) forming a wave function that is spread uniformly throughout a spherical volume up to a radius $r_0$. It is further assumed that the problem can be simplified by considering the eigenstates of the individual masses $m_i$ rather than the eigenstate of the structure as a whole. The potential energy $V(r)$ term for a small mass $m$ is then $-\frac{GMm}{r}$ outside the sphere and $\frac{GMm}{2r_0}\left(r^2 - 3r_0^{\,2}\right)$ inside the sphere. The inside term is like a 3D harmonic oscillator and by a suitable change of baseline, the Schrodinger equation may be written as $-\frac{\hbar^2}{2m}\nabla^2\psi + \frac{1}{2}kr^2\psi = E\psi$ where $k = \frac{G\pi mM}{r_0^{\,3}}$. Assuming that one particle is placed into each state (i.e., all states are filled), the sum of all the solutions to this equation up to the highest occupied level then allows an estimate of the functional form of the particle density to be obtained. Although the density of matter for all states turns out not to be constant across the sphere, it is taken this way as a first approximation in order to investigate how the density of matter changes as more states are added. For large $n$ the total number of states $n_t$ is 1/6 $n_{\max}{}^3$ where $n_{\max}$ is the principal quantum number of the highest occupied state. The effective radius $r_0$ of the structure (obtained by summing the probability distributions of all particles) can be reasonably approximated as $r_0 \approx (2n_{\max}\hbar)^{1/2}(mk)^{1/4}$ (from properties of the Laguerre polynomials). Using the formula for $k$ above with $M = n_t\, m$ then gives the total number of states (all occupied) for a radius $r_0$ as $n_t \approx 2304\,\hbar^2\left(G^3 m^9 r_0^{\,3}\right)^{-1}$. The fact that as the radius increases the number of states decreases comes about because of the change in shape of the potential and the condition that all states are filled. For a particle density of $10^5$ m$^{-3}$, it can be deduced that with all states filled, the structure has a radius of about $10^{11}$ m and a total mass of about $10^{11}$ kg but the depth of the potential well is far too small (~ $10^{-17}$ eV) for the particles to be stable. There are several ways around this difficulty without resorting to a massive central force. One alternative is to reduce the occupancy level from 100% to some small fraction, as in the galactic halo model. The various parameters are intricately connected, but this has the effect of being able to produce well-bound states that have densities of $10^5$ $m^{-3}$ and total masses of the same order as the galactic mass and should be possible because of the long radiative lifetimes. Likewise, if the mass of the individual particles is reduced substantially (~$10^{-36}$ kg), the eigenstructure can remain filled and possess densities and binding energies are in keeping with observations. An alternative, somewhat more radical hypothesis is to reduce the size of the structure. This has the effect of producing very high-density eigenstructures (~ $10^3$ kg $m^{-3}$ or greater). Whether stationary structures can exist at such densities is yet another conjecture. The matter forming the individual eigenstates could be a mixture of any stable (presumably fermionic) particles that maintained net electrical neutrality. Such eigenstructures might contribute substantially to the extra galactic dark matter and the closure of the universe. All the arguments applied to the galactic halo in the previous section (I) are valid here also. If large enough and dense enough structures existed then the potential exists for the eigenstate particles to be present in the form of atoms and they may be detectable by the atomic absorption spectra that would be obtained as the light from more distant sources shone through them.

## 3. Discussion

The concept of charged or neutral particles in the form of gravitational eigenstructures differs from that of classical cloud of hydrogen atoms or plasma in several ways that are worth repeating here:

Firstly although particles in a gas cloud might be considered to have some semi-definite energy, they are not, even approximately, in eigenstates. One might anticipate that although particles in a gas cloud will always exhibit more or less dispersed wave functions, their probability distributions are certainly not stationary. The individual gravitational eigenstates however have very precise stationary probability distributions that are not localised any more than the electrons in an atom are localised within it.

Secondly, the ability of an eigenstructure to condense or collapse into a denser object such as a star or galactic nucleus relies on either the availability of empty lower energy states with high probability transition coefficients, or a high probability of transfer to mixed, “non-eigenstates”. This collapse depends on the electrical interactions that hold condensed matter together. The crucial point is that by choosing the appropriate total mass, gravitational eigenstructures can be formed where the gravitational interaction is sufficiently large to form a totally bound and coherent structure, yet the equivalent particle densities be sufficiently low that no significant collapse involving the traditional electrical interactions of condensed matter can occur on time scales appropriate to that of the universe.

Thirdly, because the eigenstate particles are bound to the total structure as a coherent whole, their behaviour in many interactions (elastic scattering, for example) can be quite different to that of the equivalent free particle.

Lastly, because of the stationary nature of the wave functions making up the eigenstructure, charged particles in eigenstate will not radiate (except via state transitions). Hot ionised gas is normally detectable through X-ray emission or the like, but even if eigenstate particles exhibit classically equivalent high enough angular velocities and accelerations to radiate classically, they will remain “X-ray dark”.
It would appear from the investigations presented in this paper that macroscopic gravitational eigenstates clearly have properties that render them excellent dark matter candidates. The total mass of the galactic halo can be easily accommodated within the framework of a suitable eigenstructure formed around the galaxy. Unlike a gas cloud of ionised or neutral hydrogen, such a structure would be both stable with respect to gravitational collapse and largely invisible. If sufficient neutral hydrogen is present in the eigenstructure, it might be detectable through its absorption lines or recombination radiation following re-ionisation. Similar extra galactic structures with no visible component consisting of totally or partly filled gravitational eigenstates are also theoretically possible and could account for some or all of the extra galactic dark matter. Indeed the observation of the absorption spectra of light from distant quasars reveals that it has passed through many separate regions of neutral hydrogen travelling at a variety of speeds. It is suggested that if these hypothetical structures have been present from an early time in the history of the universe, a considerable amount of matter might have existed in this form and that the very large structures developed visible components while the smaller ones have remained as dark structures.

Whilst the gravitational macroscopic eigenstructure hypothesis for dark matter can explain several issues, there are clearly also many difficulties. If galactic halo eigenstructures consist of predominantly protons and electrons then where did these particles come from? Recent observation of scattered light from quasars suggests that, at 1 billion years after the Big Bang, neutral hydrogen formed at the decoupling era was re-ionised to the extent of 99.99% although the mechanism for this re-ionisation is not clear (Haiman and Loeb 1997). Either the eigenstates existed much earlier and retained their identity or there were processes at this time or earlier to ionise almost all the neutral hydrogen. A further problem concerns the level of deuterium detected in the universe. This suggests that there is still more dark matter in the form of more exotic particles. If however it becomes apparent that the types of structures discussed in this paper do exist and have been intimately connected with the expansion process since its inception, then it may be necessary to re-examined the standard model to investigate the effects that inclusion of stationary states might have on the density distribution and, in particular, on the processes occurring during the nucleosynthesis era.

Gravitational eigenstructures might be speculatively used to explain two other significant problems in astronomy, the origin of high-energy cosmic rays and the large-scale structure of the universe.

*(i) High Energy Cosmic Rays*

A significant problem in astronomy concerns the observation of numbers of very high-energy cosmic rays. The problem centres on the development of a suitable mechanism for the production of these since at present, no known physical process can account for the very high energies observed. The production of such cosmic rays follows quite naturally however from level decay within eigenstructures that possess a massive central core. In traditional atomic physics the energy release accompanying electron demotion within the atom is of the order of at most tens of electron volts and electromagnetic radiation is the only possible type of emission. Conservation laws could also be satisfied with the emission of electron-positron or proton-antiproton pairs instead of photons. The energy changes for the inner states near a massive but small central potential provide potentially vast amounts of surplus energy to be carried away as kinetic energy of the particles produced and one might expect to see such transitions if the lifetimes of these inner states are sufficiently short. Although such transitions would probably have taken place some time ago in the Milky Way, they may be observable in young objects such as quasars.

*(ii) Large Scale Structure*

Another speculative explanation of the large-scale structure of the universe may be made using macro-eigenstates. When a plate covered with fine sand is caused to resonate by rubbing its edge with say, a violin bow, standing waves are set up and sand collects in the nodes of those waves. If during inflation, when light was able to travel many times the size of the visible universe, standing waves were set up, then the potential exists for an equivalent three-dimensional array of walls of collecting matter to form. If the standing waves were electromagnetic or gravitational in origin the matter would tend to preferentially collect at the nodal surfaces. Alternatively matter eigenstates may have formed across the entire universe with the consequence that matter would manifest itself in regions where the probability density was greatest, nevertheless again in sheet-like areas.

**4. Conclusion**

This paper has not attempted to prove or disprove the existence of macroscopic gravitational eigenstructures but rather to show, by looking at macroscopic material in the universe from a quantum perspective, (1) that there is very little in traditional physical theory that forbids the existence of eigenstructures and (2) that they do provide a possible (and perhaps reasonable) explanation of the composition of dark matter and some other cosmological phenomena.

Although the models used have involved extremely gross approximations, the broad conclusions from these models should remain generally valid when more realistic details are included, and suggest that gravitational macroscopic eigenstates can be used to explain some fundamental problems in modern cosmology. Results also suggest that it would be worth re-examining Big Bang models with the inclusion of macroscopic stationary eigenstates at all times in the past

---

## Appendix A. Expansion rate of the free particle wave function

It is a direct prediction of quantum theory that the spatial spread of the wave function will always increase with time (both in the forward and backward direction) for a free, non-interacting particle.

To calculate volume changes a three-dimensional treatment of the free particle is required. An initial ($t$=0) free particle wave function may be easily written down as any relatively localised function in $x$ and $p$ space. Standard texts (e.g. (Schiff 1965)) use a Gaussian functional form which gives the *minimal uncertainty* one-dimensional wave packet, that is,

$$\psi(x) = (2\pi(\Delta x)^2)^{-\frac{1}{4}} \exp[-\frac{(x-\langle x\rangle)^2}{4(\Delta x)^2} + \frac{i\langle p\rangle x}{\hbar}], \tag{A1}$$

where the initial uncertainties in $x$ and $p$ are given by $(\Delta x)^2 = \left\langle (x - \langle x \rangle)^2 \right\rangle$ and $(\Delta p)^2 = \left\langle (p - \langle p \rangle)^2 \right\rangle$ (ie. on the basis that the uncertainties are given as the root mean square deviation from the expectation values). Although the derivation of the minimal form is rather tedious, it can be straightforwardly extended to three dimensions to give

$$\psi(x_j) = (8\pi^3 (\prod_j \Delta x_j)^2)^{-\frac{1}{4}} \exp[-\sum_j \frac{(x_j - \langle x_j \rangle)^2}{4(\Delta x_j)^2} + \frac{i}{\hbar} \sum_j \langle p_j \rangle x_j] \qquad \text{(A2)}$$

where $\psi(\mathbf{r}) = \psi(x_j)$ and $j$ = 1 to 3 gives the three spatial dimensions.

The generalised time dependant solution to (A1) is then obtained by writing $\psi(\mathbf{r},t) = \psi(x_j,t)$ as the integral summation of momentum eigenfunctions $u_k(x_j)$:

$$\psi(x_j,t) = \iiint A_{k_j} u_k(x_j) \exp[-iE_k t/\hbar] dk_j \qquad \text{(A3)}$$

where $u_k(x_j) = u_k(\mathbf{r}) = \left(8\pi^3\right)^{-\frac{1}{2}} \exp(i\mathbf{k}\cdot\mathbf{r})$, $E_k = \frac{\hbar^2 k^2}{2m} = \frac{\hbar^2}{2m} \sum_j k_j^{\,2}$ = the energy of the eigenstate $u_k(x_j)$, and the $A_{k_j}$ are to be determined from the form of (A2). After evaluating the integrals, some algebraic manipulation and using $\hbar / \Delta x_j = 2\Delta p_j$ (we can use the equality because of the specific definitions of $\Delta x_j$ and $\Delta p_j$, and their relation to the Gaussian form of (A2)) we get

$$\psi(x_j,t) = (2\pi)^{-\frac{3}{4}} \exp\left[-\frac{iE_{\langle k\rangle}t}{\hbar}\right] \prod_j \left(\Delta x_j + \frac{i\Delta p_j t}{m}\right)^{-\frac{1}{2}} \exp\left[\frac{-\left(x_j - \langle x_j\rangle - \frac{\langle p_j\rangle t}{m}\right)^2}{4\Delta x_j\left(\Delta x_j + \frac{i\Delta p_j t}{m}\right)} + \frac{i\langle p_j\rangle x_j}{\hbar}\right]$$

(A4)

where $E_{\langle k\rangle}$ is the expectation value of the energy (kinetic for this free particle).

The three-dimensional position density from which the changes in volume may be found is then given by

$$\psi^*\psi = (2\pi)^{-\frac{3}{2}} \prod_j \left(\Delta x_j^{\,2} + \frac{\Delta p_j^{\,2} t^2}{m^2}\right)^{-\frac{1}{2}} \exp\left[\frac{-\left(x_j - \langle x_j\rangle - \frac{\langle p_j\rangle t}{m}\right)^2}{2\left(\Delta x_j^{\,2} + \frac{\Delta p_j^{\,2} t^2}{m^2}\right)}\right] \qquad \text{(A5)}$$

where again $j$ refers to the three Cartesian axes. Figure 2 shows the effective volume of a wave packet, defined as $\Delta V(t) = \prod_j \Delta x_j(t)$ where $\Delta x_j(t) = \left(\Delta x_j^{\,2} + \frac{\Delta p_j^{\,2} t^2}{m^2}\right)^{\frac{1}{2}}$, of a free particle as a function of time for different values of the initial effective volume $\Delta V_i = \prod_j \Delta x_j$.

As , these changes can be remarkably fast. An electron emitted from an interaction constricted to nuclear dimensions, say 5 fm cubed, will reach 1 litre size within 10 ps, and galactic size within about 30 Gs (1000 years). It should be noted that these quantum theory predictions represent the spread obtained using a specific form for the wave function that yields the minimum spread and that no allowance has been made for increase in wave function extent due the expansion of space, potentially significant in the early universe.

Appendix B. Approximate extent of the function $(R_{nl}r)^2$ for large *n*.

The general form of the radial wave functions can be used to obtain a measure of its extent for large *n*. An empirical observation of the Laguerre functions comes from examining the position $r_{\text{finalmax}}$ the final maximum of the radial part of probability density function $(R_{nl}r)^2$ and comparing it to the maximums of the truncated functions formed using only last one or more of the terms in the Laguerre summation. It is possible to write down explicit expressions for the maxima of these contrived functions. The various maxima in the Laguerre polynomials (and hence in the radial probability functions) arise as a result of a sensitive balance between successively higher powers of *r* and their coefficients, the final maximum being a competitive interaction between the ultimate divergence of the Laguerre summation due to the highest order terms and the factor of $e^{-\frac{\mu GmMr}{n\hbar^2}}$. One might expect therefore to see some empirical relation between the position of the final maximum in the probability density function $(R_{nl}r)^2 = \left( e^{-\frac{\mu GmMr}{n\hbar^2}} r L_{n+l}^{2l+1}\left(\frac{2\mu GmMr}{n\hbar^2}\right)\right)^2$ and the positions of the maxima of the truncated functions formed using only the last one or more terms of the Laguerre summation, that is

$$\left(R_{n\ell}r\right)^2{}_1 = \left( e^{-\frac{\mu GmMr}{n\hbar^2}} r \frac{(n+l)!\left(\frac{\mu GmMr}{n\hbar^2}\right)^{n-l-1}}{(2l+n-l)!(n-l-1)!} \right)^2,$$

$$\left(R_{n\ell}r\right)^2{}_2 = \left( e^{-\frac{\mu GmMr}{n\hbar^2}} r \sum_{k=n-l-2}^{n-l-1}(-1)^{k+2l+1} \frac{(n+l)!\left(\frac{\mu GmMr}{n\hbar^2}\right)^{k}}{(n-l-1-k)!(2l+1+k)!k!} \right)^2 \quad \text{or}$$

$$\left(R_{n\ell}r\right)^2{}_3 = \left( e^{-\frac{\mu GmMr}{n\hbar^2}} r \sum_{k=n-l-3}^{n-l-1}(-1)^{k+2l+1} \frac{(n+l)!\left(\frac{\mu GmMr}{n\hbar^2}\right)^{k}}{(n-l-1-k)!(2l+1+k)!k!} \right)^2 \quad \text{(B1)}$$

Formulae for the maxima of the truncated functions B1 are exactly determinable from the zeros their differentials and it is found that, there is a converging relationship for large $n$, between the last maxima of $(R_{nl}r)^2$ and at least one the roots of each of the differentials of B1.

Figure 3 shows how the relative percentage difference between position of the final maximum in the probability function $(R_{nl}r)^2$ and the maxima of the functions, $\left(R_{nl}r\right)^2{}_1$, $\left(R_{nl}r\right)^2{}_2$ and $\left(R_{n\ell}r\right)^2{}_3$ varies as a function of $n$ for $l = 0$.

Although all of the solutions to the maximums of the functions $\left(R_{n\ell}r\right)^2{}_1$, $\left(R_{n\ell}r\right)^2{}_2$ and $\left(R_{n\ell}r\right)^2{}_3$ lead to empirical relationships with the final maximum of the probability density function $(R_{nl}r)^2$, those shown in figure 3 lead to the simple relationship that, for $l = 0$, as $n \to \infty$,

$$r_{\text{finalmax}} \to 2n^2 \frac{\hbar^2}{\mu GmM} = 2n^2 b_0 , \tag{B2}$$

where $r_{\text{finalmax}}$ is the position of the final maximum of the probability density function $(R_{nl}r)^2$ and $n$ is the principal quantum number of the eigenstate. It is possible to have an estimate of both the size and shape of the eigenstates for large $n$ ($l$=0) without specifically calculating equation (4). For ($l = n$-1), the position of the $r_{\text{finalmax}}$ ($\approx n^2 b_0$) value is directly calculable. Consideration of $(R_{nl}r)^2$ for ($l = n$-2), ($l = n$-3), etc., verifies that as $l$ decreases relative to $n$, successive $(R_{nl}r)^2$ maxima spread out in an approximate square root dependence around this peak.

Appendix C. Procedure for calculation of some *TR* values using matrix elements $p_{if}$ for large $n$ and $l$

The matrix element $p_{if}$ can be explicitly written in its components, $p_{ifx}$ , $p_{ify}$ and $p_{ifz}$ by

$$p_{if} = \sqrt{p_{ifx}{}^2 + p_{ify}{}^2 + p_{ifz}{}^2} \tag{C1}$$

where

$$p_{ifx} = \int_0^\infty \int_0^\pi \int_0^{2\pi} R^*_{nf,lf}(r) Y^*_{lf,mf} er R_{ni,li}(r) Y_{li,mi} r^2 \sin(\theta)\cos(\phi)\sin(\theta) d\phi d\theta\, dr \tag{C2}$$

$$p_{ify} = \int_0^{\infty}\int_0^{\pi}\int_0^{2\pi} R^*_{nf,lf}(r)Y^*_{lf,mf}erR_{ni,li}(r)Y_{li,mi}r^2 \sin(\theta)\sin(\phi)\sin(\theta)d\phi d\theta\, dr \tag{C3}$$

$$p_{ifz} = \int_0^{\infty}\int_0^{\pi}\int_0^{2\pi} R^*_{nf,lf}(r)Y^*_{lf,mf}erR_{ni,li}(r)Y_{li,mi}r^2 \cos(\theta)\sin(\theta)d\phi d\theta\, dr \tag{C4}$$

The total decay rate for any level will be sum of the transition rates through each of its available decay channels. Finding these transition rates requires calculation of the relevant values of ω and $p_{if}$, the latter being obtained using the initial and final eigenstate solutions to equation (1) (with *n* values up to $10^{30}$ or more) and incorporating them into equations (C1) to (C4).

There are many more states available than is required to provide the amount of dark matter estimated to be present in the galactic halo. As noted however, it is impossible to write down explicit forms for these eigenfunctions in general, because of the huge number of terms involved when *n* is large (equation (4)). It is possible however to obtain decay rates for some specific cases. Fortunately by analogy with atomic systems, it might be expected that the longer-lived states are the high angular momentum, large principal quantum number states (large *n* and *l*). When *l* is very close to *n* (say $l = n - a$ where *a* is 1, 2, 3 ...), it is easy to write down $R_{nl}(r)$ because there are only a small number of terms in the summation ($k = 0$ to $k = a - 1$). A further simplification arises through the explicit calculations of the angular dependency for which it is easily shown that, when $\Delta m = \pm 1$, $p_{ifz}$ is zero and when $\Delta m = 0$, $p_{ifx}$ and $p_{ify}$ are zero.

In figure 1 each point represents the $2\,l + 1$ sublevels corresponding to the $2\,l + 1$ possible z projections of the angular momentum state available. The standard dipole selection rules apply for transitions between states, that is $\Delta l = \pm 1$ and $\Delta m = 0$ or $\pm 1$. Thus in this diagram downward transitions originating from states such as A along the first diagonal ($l = n$-1), can decay via a change in $n$ of only 1 per transition because of the selection rule $\Delta l = \pm 1$. For transitions originating from states such as B along the next diagonal ($l = n$-2), two decay channels are possible, with $\Delta n = 1$ or 2. Transitions originating from states like C and D further to the right in the figure ($l = n$-3, $n$-4, $n$-5 etc.) have successively more decay channels available with larger $\Delta n$ values. The significant point to note here however is that each time a transition from an initial state ($l_i$, $n_i$) to a final state ($l_f$, $n_f$) occurs, $n_f - l_f \le n_i - l_i$. No transitions occur that ever take a state further from the ($l = n$-1) diagonal. The trend therefore is that, as radiative decay proceeds, there is a net migration of states towards the ($l = n$-1) diagonal at the left of the diagram, that is, to *relatively* higher $l$ compared to $n$ values (eg E $\rightarrow$ F $\rightarrow$ G $\rightarrow$ H).

Each component of $p_{if}$ may be split into the separate radial and angular integrals. Eg

$$p_{ifz} = \int_0^\infty eR^*_{nf,lf}(r)r^3 R_{ni,li}(r)\,dr \int_0^\pi \int_0^{2\pi} Y^*_{lf,mf} Y_{li,mi} \cos(\theta)\sin(\theta)\,d\phi\, d\theta \,.$$

The angular components, $\int_0^\pi \int_0^{2\pi} Y^*_{lf,mf} Y_{li,mi} \sin(\theta)\cos(\phi)\sin(\theta)\,d\phi\, d\theta$,

$\int_0^\pi \int_0^{2\pi} Y^*_{lf,mf} Y_{li,mi} \sin(\theta)\sin(\phi)\sin(\theta)\,d\phi\, d\theta$ and $\int_0^\pi \int_0^{2\pi} Y^*_{lf,mf} Y_{li,mi} \cos(\theta)\sin(\theta)\,d\phi\, d\theta$

corresponding to $p_{ifx}, p_{ify}$ and $p_{ifz}$, respectively, depend on the initial and final values of $m$. For example when the initial, $m_i$, and final, $m_f$, values of $m$ are both $l - 1$, so that $\Delta m = 0$ then the total angular component of $p_{if}$ is just that for $p_{ifz}$. For simplicity only the cases where $m$ is even are presented here (similar results apply for $m$ odd).

$Y_{l,m}$ may be written as $Y_{l,m} = \sqrt{\frac{2l+1}{2}\frac{(l-m)!}{(l+m)!}}P_l^m(\cos\theta)$ (Arfken and Weber 1995).

Substituting $m = l - j$, using Rodrigues' formula and the binomial theorem, and adjusting the summation limits appropriately, gives the initial state $Y_{li,mi} = Y_{l,m}$ as

$$
\begin{aligned}
Y_{l,m} &= \sqrt{\frac{2l+1}{2}\frac{(l-m)!}{(l+m)!}}P_l^m(\cos\theta) \\
&= (-1)^m \sqrt{\frac{2l+1}{4\pi}\frac{(l-m)!}{(l+m)!}}\exp(im\phi)\frac{1}{2^l l!}(1-\cos^2\theta)^{m/2}\frac{\partial^{l+m}}{\partial(\cos\theta)^{l+m}}(\cos^2\theta-1)^l \\
&= (-1)^{(l-j)}\sqrt{\frac{2l+1}{4\pi}\frac{j!}{(2l-j)!}}\exp(i(l-j)\phi)\frac{1}{2^l l!}(1-\cos^2\theta)^{(l-j)/2} \times \\
&\left\{\sum_{k=l-j/2}^{l}\left[(-1)^{(l-k)}\frac{(2k)!(\cos\theta)^{(2k-2l+j)}}{(2k-2l+j)!}\frac{l!}{(l-k)!k!}\right]\right\}
\end{aligned}
\tag{C5}
$$

The final state $Y_{lf,mf} = Y_{l-1,m}$ has $m = (l-1)-(j-1)$ and is likewise given by

$$
\begin{aligned}
&= (-1)^{(l-j)}\sqrt{\frac{2l-1}{4\pi}\frac{(l+j)!}{(2l-j-1)!}}\exp(i(l-j)\phi)\frac{1}{2^{(l-1)}(l-1)!}(1-\cos^2\theta)^{(l-j)/2} \times \\
&\left\{\sum_{k=l-j/2}^{(l-1)}\left[(-1)^{(l-k-1)}\frac{(2k)!(\cos\theta)^{(2k-2l+j+1)}}{(2k+j)!}\frac{(l-1)!}{(l-k-1)!k!}\right]\right\}
\end{aligned}
\tag{C6}
$$

The correctness of the above formula may be easily verified by substitution of some specific values for *l* and *m*. The angular part of $I_z$ then becomes

$$\int_0^{2\pi}\int_0^{\pi} \begin{matrix} (-1)^{(l-j)}\sqrt{\frac{2l+1}{4\pi}\frac{j!}{(2l-j)!}}\exp(i(l-j)\phi)\frac{1}{2^l l!}(1-\cos^2\theta)^{(l-j)/2}\times \\ (-1)^{(l-j)}\sqrt{\frac{2l-1}{4\pi}\frac{(l+j)!}{(2l-j-1)!}}\exp(i(l-j)\phi)\frac{1}{2^{(l-1)}(l-1)!}(1-\cos^2\theta)^{(l-j)/2}\times \\ \left\{\sum_{k=l-j/2}^{l}\left[(-1)^{(l-k)}\frac{(2k)!(\cos\theta)^{(2k-2l+j)}}{(2k-2l+j)!}\frac{l!}{(l-k)!k!}\right]\right\}\times \\ \left\{\sum_{k=l-j/2}^{(l-1)}\left[(-1)^{(l-k-1)}\frac{(2k)!(\cos\theta)^{(2k-2l+j+1)}}{(2k+j)!}\frac{(l-1)!}{(l-k-1)!k!}\right]\right\}\cos\theta\sin\theta \end{matrix} d\theta d\phi \tag{C7}$$

Integrating over $\phi$ and putting

$P_p = \frac{(-1)^p(-1)^{j/2}l!(2p-j+2l)!}{(p-j/2+l)!(j/2-p)!(2p)!}$ and $Q_q = \frac{(-1)^{q+1}(-1)^{j/2}l!(2p-j+2l)!}{(q-j/2+l)!(j/2-q-1)!(2q+1)!}$ gives,

again after adjusting limits, $I_z$ as

$$\begin{matrix} \frac{1}{2^{2l}l!(l-1)!}\left(\sqrt{\frac{(4l^2-1)j!(j-1)!}{(2j-1)!(2l-j-1)!}}\right)\times \\ \int_0^{\pi}\cos\theta\sin\theta^{(2(l-j)+1)}\left(\sum_{p=0}^{j/2}P_p\cos^{2p}\theta\right)\left(\sum_{p=0}^{j/2-1}Q_q\cos^{2q+1}\theta\right)d\theta \end{matrix} \tag{C8}$$

On expanding the product of the summations and integrating this expression may be written as

$$\frac{1}{2^{2l}l!(l-1)!}\left(\sqrt{\frac{(4l^2-1)j!(j-1)!}{(2j-1)!(2l-j-1)!}}\right)\times$$
$$\left(\left(\sum_{\xi=0}^{j/2-1}\frac{2^{2(l-j)}(2\xi+2)!(l-j)!(2l-2j+1)!\left(l-j+\xi+1)!\right)}{(l-j+1/2)(2l-2j)!(\xi+1)!(2l-2j+2\xi+3)!}\left(\sum_{i=0}^{\xi}P_iQ_{\xi-i}\right)\right)+\left(\sum_{\xi=j/2}^{j-1}\frac{2^{2(l-j)}(2\xi+2)!(l-j)!(2l-2j+1)!\left(l-j+\xi+1)!\right)}{(l-j+1/2)(2l-2j)!(\xi+1)!(2l-2j+2\xi+3)!}\left(\sum_{i=\xi-j/2+1}^{j/2}P_iQ_{\xi-i}\right)\right)\right) \quad \text{(C9)}$$

This gives the final value of $I_z$ as

$$\frac{1}{2^{2l}l!(l-1)!}\left(\sqrt{\frac{(4l^2-1)j!(j-1)!}{(2j-1)!(2l-j-1)!}}\right)\times$$
$$\left(\left(\sum_{\xi=0}^{j/2-1}\left(\frac{2^{2(l-j)}(2\xi+2)!(l-j)!(2l-2j+1)!\left(l-j+\xi+1)!\right)}{(l-j+1/2)(2l-2j)!(\xi+1)!(2l-2j+2\xi+3)!}\times\left(\sum_{i=0}^{\xi}\left(\frac{(-1)^{\xi+1}l!(2i-j+2l)!(l-1)!(2\xi-2i-j+2l)!}{(i-j/2+l)!(2l-2j)!(2i)!(\xi-i-j/2+l)!(j/2+i-\xi-1)!(2\xi-2i+1)!}\right)\right)\right)\right)+\right.$$
$$\left.\left(\sum_{\xi=j/2}^{j-1}\left(\frac{2^{2(l-j)}(2\xi+2)!(l-j)!(2l-2j+1)!\left(l-j+\xi+1)!\right)}{(l-j+1/2)(2l-2j)!(\xi+1)!(2l-2j+2\xi+3)!}\times\left(\sum_{i=\xi-j/2+1}^{j/2}\left(\frac{(-1)^{\xi+1}l!(2i-j+2l)!(l-1)!(2\xi-2i-j+2l)!}{(i-j/2+l)!(2l-2j)!(2i)!(\xi-i-j/2+l)!(j/2+i-\xi-1)!(2\xi-2i+1)!}\right)\right)\right)\right)\right)$$

(C10)

It can be shown that the limiting value of this expression for large $l$ is $1/\sqrt{2l+1}$, provided that $l$ is large compared to j. In the situation considered in this paper, values of $l$ are generally close to $n$, which is also very large. Since $j=n-l$, $j$ is small relative to $l$ and this condition is easily satisfied.

When $\Delta m = \pm 1$ the integral components $I_x = \int_0^{\pi}\int_0^{2\pi} Y^*_{lf,mf} Y_{li,mi} \sin(\theta)\sin(\phi)\sin(\theta) d\phi d\theta$ and $I_y = \int_0^{\pi}\int_0^{2\pi} Y^*_{lf,mf} Y_{li,mi} \cos(\theta)\sin(\theta) d\phi d\theta$ are non-zero. In a similar way to the above treatment, it can be shown that these integrals are small ($\leq \pi/2$) under the same conditions.

A general form for the radial component of the overlap integral $\int_0^{\infty} eR^*_{nf,lf}(r) r^3 R_{ni,li}(r) r^2\, dr$ is given below, although estimates of its behaviour are generally less predictable than those for the angular component and usually require direct computation of each specific case. Furthermore, because of computing limitations (for example calculation of $10^{30}$ factorial), the actual numerical calculations were undertaken using the exponent of the logarithm of each result, and require the use of Stirling's approximation for the factorial function.

The radial component of the initial state is taken as $(n_i, l_i) = (n_i, l) = (n_i, n_i - j)$ where $j = n_i - l$. The final state $(n_f, l_f)$ requires that $l_f = l - 1$ or $l_f = l + 1$ and is taken here as $(n_f, l_f) = (n_f, l - 1) = (n_f, n_f - j - 1)$ (with a similar result for $l_f = l + 1$). (Note also that the parameters $P_p$, $Q_q$ and $j$ used below are not the same as those used in the angular integrals above.) Introducing the constant $b_0 = \frac{\hbar^2}{\mu G m M}$, the initial and final states written in terms of $j$ therefore respectively become, using equation 4

$R_{n_i,l} =$

$$\left(\left(\frac{2}{n_i b_0}\right)^3\left(\frac{(j-1)!}{2n_i((2n_i-j)!)^3}\right)\right)^{\frac{1}{2}}\exp\left(-\frac{r}{n_i b_0}\right)\left(\frac{2r}{n_i b_0}\right)^{(n_i-j)}\left(\sum_{k=0}^{(j-1)}(-1)^k\frac{((2n_i-j)!)^2\left(\frac{2r}{n_i b_0}\right)^k}{(j-k-1)!(2n_i-2j+k+1)!k!}\right)$$

and

$$R_{n_f,l-1} = \left(\left(\frac{2}{n_i b_0}\right)^3\left(\frac{(n_f-n_i+j-2)!}{2n_f((n_i+n_f-j-1)!)^3}\right)\right)^{\frac{1}{2}}\exp\left(-\frac{r}{n_i b_0}\right)\left(\frac{2r}{n_i b_0}\right)^{(n_i-j-1)}\times$$
$$\left(\sum_{k=0}^{j}(-1)^k\frac{((n_i+n_f-j-1)!)^2\left(\frac{2r}{n_i b_0}\right)^k}{(n_f-n_i+j-k)!(2n_i-2j+k-1)!k!}\right) \quad \text{(C11)}$$

The radial integral $\int_0^\infty eR^*_{nf,lf}(r)r^3R_{ni,li}(r)r^2\,dr$ becomes

$$\frac{e}{2}\int_0^\infty\left[\begin{array}{l}\left(\frac{(j-1)!(n_f-n_i+j)!}{n_i n_f((2n_i-j)!)^3((n_f+n_i-j-1)!)^3}\right)^{\frac{1}{2}} n_i^{\left(-\frac{3}{2}+j-n_i\right)} n_f^{\left(-\frac{1}{2}+j-n_i\right)}\exp\left(-\left(\frac{n_f+n_i}{n_i n_f b_0}\right)r\right)\times \\ \left(\frac{2r}{b_0}\right)^{(2n_i-2j+2)}\left(\sum_{p=0}^{(j-1)}P_p\left(\frac{2r}{b_0}\right)^p\right)\left(\sum_{q=0}^{j}Q_q\left(\frac{2r}{b_0}\right)^q\right)\end{array}\right]dr$$

(C12)

where

$$P_p = (-1)^p \frac{((2n_i - j)!)^2}{n_i{}^p (j-p-1)!(2n_i - 2j + p + 1)!\,p!}$$ and

$$Q_q = (-1)^q \frac{((n_i + n_f - j - 1)!)^2}{n_i{}^q (j - q + n_f - n_i)!(2n_i - 2j + q - +1)!\,q!}.$$

The product of summations over $p$ and $q$ can be then expanded to give a new summation over $\xi$ and $s$ which is directly integrable:

$$\frac{e}{2}\int_0^\infty \left[ \begin{array}{l} \left( \dfrac{(j-1)!(n_f - n_i + j)!}{n_i n_f ((2n_i - j)!)^3 ((n_f + n_i - j - 1)!)^3} \right)^{\frac{1}{2}} n_i^{\left(-\frac{3}{2}+j-n_i\right)} n_f^{\left(-\frac{1}{2}+j-n_i\right)} \times \\ \left[ \begin{array}{l} \left\{ \displaystyle\sum_{\xi=0}^{(j-1)} \left( \sum_{s=0}^{\xi} P_s Q_{\xi-s} \right) \exp\left( -\left( \frac{n_f + n_i}{n_i n_f b_0} \right) r \right) \left( \frac{2r}{b_0} \right)^{\xi+2n_i-2j+2} \right\} + \\ \left\{ \displaystyle\sum_{\xi=j}^{(2j-1)} \left( \sum_{s=\xi-j}^{j-1} P_s Q_{\xi-s} \right) \exp\left( -\left( \frac{n_f + n_i}{n_i n_f b_0} \right) r \right) \left( \frac{2r}{b_0} \right)^{\xi+2n_i-2j+2} \right\} \end{array} \right] \end{array} \right] dr \qquad \text{(C13)}$$

Using the fact that

$$\int_0^\infty \exp\left( -\left( \frac{n_f + n_i}{n_i n_f b_0} \right) r \right) \left( \frac{2r}{b_0} \right)^{\xi+2n_i-2j+2} dr = \frac{b_0 (\xi + 2n_i - 2j + 2)!}{2} \left( \frac{2n_i n_f}{(n_i + n_f)} \right)^{(\xi+2n_i-2j+3)}$$

gives $\int_0^\infty e R^*_{nf,lf}(r) r^3 R_{ni,li}(r) r^2\, dr$ as

$$\frac{e}{2}\left[\left(\frac{(j-1)!(n_f-n_i+j)!}{n_i n_f((2n_i-j)!)^3((n_f+n_i-j-1)!)^3}\right)^{\frac{1}{2}} n_i^{\left(-\frac{3}{2}+j-n_i\right)} n_f^{\left(-\frac{1}{2}+j-n_i\right)} \times \left[\left\{\sum_{\xi=0}^{(j-1)}\left(\sum_{s=0}^{\xi}P_sQ_{\xi-s}\right)\frac{b_0(\xi+2n_i-2j+2)!}{2}\left(\frac{2n_in_f}{(n_i+n_f)}\right)^{(\xi+2n_i-2j+3)}\right\}+\left\{\sum_{\xi=j}^{(2j-1)}\left(\sum_{s=\xi-j}^{j-1}P_sQ_{\xi-s}\right)\frac{b_0(\xi+2n_i-2j+2)!}{2}\left(\frac{2n_in_f}{(n_i+n_f)}\right)^{(\xi+2n_i-2j+3)}\right\}\right]\right] \qquad \text{(C14)}$$

which on substitution for $P_p$ and $Q_q$ gives:

$$\frac{e}{2}\left[\left(\frac{(j-1)!(n_f-n_i+j)!}{n_i n_f((2n_i-j)!)^3((n_f+n_i-j-1)!)^3}\right)^{\frac{1}{2}} n_i^{\left(-\frac{3}{2}+j-n_i\right)} n_f^{\left(-\frac{1}{2}+j-n_i\right)} \times \left[\left\{\sum_{\xi=0}^{(j-1)}\left(\sum_{s=0}^{\xi}(-1)^{\xi}\frac{((2n_i-j)!(n_i+n_f-j-1)!)^2}{n_i^{\xi}(j-s-1)!(j-(\xi-s)+n_f-n_i)!(2n_i-2j+(\xi-s)-+1)!(\xi-s)!(2n_i-2j+s+1)!s!}\right)\frac{b_0(\xi+2n_i-2j+2)!}{2}\left(\frac{2n_in_f}{(n_i+n_f)}\right)^{(\xi+2n_i-2j+3)}\right\}+\left\{\sum_{\xi=j}^{(2j-1)}\left(\sum_{s=\xi-j}^{j-1}(-1)^{\xi}\frac{((2n_i-j)!(n_i+n_f-j-1)!)^2}{n_i^{\xi}(j-s-1)!(j-(\xi-s)+n_f-n_i)!(2n_i-2j+(\xi-s)-+1)!(\xi-s)!(2n_i-2j+s+1)!s!}\right)\frac{b_0(\xi+2n_i-2j+2)!}{2}\left(\frac{2n_in_f}{(n_i+n_f)}\right)^{(\xi+2n_i-2j+3)}\right\}\right]\right]$$

(C15)

This expression now gives the radial component of the overlap integral

$\int_0^{\infty} eR^{*}_{nf,lf}(r)r^3R_{ni,li}(r)r^2\,dr$, for any general values of $n_i, n_f$ and $j(= n_i - l)$. For example

substituting $n_f = n_i - 1$ and $j = 1$ into equation (C9) gives an explicit expression for transitions of the type A to A′ ( $(n_i = n_i, l = n_i - 1) \rightarrow (n_f = n_i - 1, l = n_i - 2)$ ) shown in figure 1 as:

$$\int_0^\infty eR^*_{nf,lf}(r)r^3 R_{ni,li}(r)r^2\,dr \;=\; 2^{2n_i} eb_0 \left( \frac{n_i(n_i-1)}{(2n_i-1)^2} \right)^{n_i+1} \sqrt{(2n_i-1)^3(2n_i-2)} \qquad \text{(C16)}$$

which, for large $n_i$ becomes $eb_0 n_i^2$.

Substituting the relevant quantities into equations (C1) to (C4) and (9) then gives the transition rate of the eigenstate. As explained in the text, this results in lifetimes that are far older than the age of the universe.

Eigenstates on the ($l = n - j$) th diagonal have radial eigenfunctions that are Laguerre polynomials with $j$ turning points. Consider now transitions of the type ($(n_i = n_i, l_i = n_i - 2) \rightarrow (n_f = n_i - 1, l_f = n_i - 3)$ shown in figure 1. These transitions take place between two '2- turning point' Laguerre functions, and $n_f = n_i - 1$ and $j = 2$. In this case equation (C15) yields a summation involving with four terms, each of which involves several factorial functions. These may be reduced to the square root of products of terms, which in turn may be simplified using a Taylor expansion to second order around $n_i = 0$. The result gives $eb_0 n_i^2$ for large $n_i$. The radiative decay time is therefore essentially the same as that for transitions of the type A to A′. It can be shown with some difficulty that, whenever transitions take place between two '$j$-turning point' Laguerre eigenfunctions, that is, along the same diagonal so that $n \rightarrow n-1$ and $l \rightarrow l-1$, then the decay rate is $eb_0 n_i^2$ provided $n >> j$.

Transitions like B to B″ in figure 1 involve overlap integrals where the wave function's radial components have very different shapes (B is a 1-turning point function while B″ is a 2-turning point function). As a result, it would be expected in this case that the radial part of the overlap integral would be much smaller than either type A to A′ or B to B′ transitions. Equation (C15) gives

$$\int_0^\infty eR^*_{nf,lf}(r)r^3R_{ni,li}(r)r^2\,dr \;=\; 2^{2n_i-2}eb_0\left(\frac{n_i(n_i-2)}{(2n_i-2)^2}\right)^{n_i}\sqrt{(2n_i-2)(2n_i-3)(2n_i-4)}, \quad \text{(C17)}$$

which reduces to $\frac{eb_0}{\sqrt{2}}n_i^{\frac{3}{2}}$ again provided $n_i >> j$.

A general formula for the value of $\int_0^\infty eR^*_{nf,lf}(r)r^3R_{ni,li}(r)r^2\,dr$ may be obtained for any transition like E to E′ of figure 1, which originates from an arbitrary *j*-turning point radial Laguerre polynomial state $(n_i, l_i = n_i - j)$ and ends on a 1-turning point state ($n_f = n_i - j, l_f = n_f - 1$, lying on the left diagonal). It may be shown that in this case equation (C15) reduces to

$$\frac{e}{2}\left[\left(\frac{(j-1)!(n_f-n_i+j)!}{n_in_f((2n_i-j)!)^3((n_f+n_i-j-1)!)^3}\right)^{\frac{1}{2}} n_i^{\left(-\frac{3}{2}+j-n_i\right)} n_f^{\left(-\frac{1}{2}+j-n_i\right)} \times \left[\left\{\begin{array}{l}\displaystyle\sum_{\xi=0}^{(j-1)}\left(\sum_{s=0}^{\xi}(-1)^{\xi}\frac{((2n_i-j)!(n_i+n_f-j-1)!)^2}{n_i^{\xi}(j-s-1)!(j-(\xi-s)+n_f-n_i)!(2n_i-2j+(\xi-s)-+1)!(\xi-s)!(2n_i-2j+s+1)!s!}\right)\\ \displaystyle\frac{b_0(\xi+2n_i-2j+2)!}{2}\left(\frac{2n_in_f}{(n_i+n_f)}\right)^{(\xi+2n_i-2j+3)}\end{array}\right\}\right]\right]$$

(C18)

After some algebraic manipulation, the double summation over $\xi$ and $s$ may be simplified and equation (C18) becomes

$$\frac{eb_0}{2}\sqrt{(j-1)!}\left(\frac{n_i n_f}{(n_i+n_f)^2}\right)^{(n_f+1)} 2^{2n_i-1}\left(\prod_{i=0}^{j}(n_i+n_f-i)^{\frac{1}{2}}\right)\left(\sum_{i=1}^{j}\left((-1)^{i-1}\left(\frac{n_f}{(n_i+n_f)}\right)^i \frac{2^{i-2j+2}(2n_f+i+1)}{(i-1)!(j-i)!}\right)\right)$$

(C19)

Carrying out the summation over $i$ and simplifying the product, (C19) becomes

$$eb_0\left(\frac{4n_i n_f}{(n_i+n_f)^2}\right)^{(n_f+1)} \frac{j^{j-2}n_i n_f}{(n_i+n_f)^j}\sqrt{\frac{(n_i+n_f)!}{(j-1)!(2n_f-1)!}} \qquad \text{(C20)}$$

Using Stirling's approximation (C20) may be written as

$$eb_0 n_i\left(\frac{4n_i n_f}{(n_i+n_f)^2}\right)^{(n_f+1)}\left(\frac{n_i+n_f+1}{2n_f}\right)^{n_i}\left(\frac{(n_i+n_f+1)n_f^{\ 5}}{\pi e^2 j^7}\right)^{\frac{1}{4}}\sqrt{\left(\frac{4n_f^{\ 2}j}{(n_i+n_f+1)(n_i+n_f)^2}\right)^j}$$

(C21)

Provided $n_i >> j$, then $n_i \approx n_f (= n_i - j)$, and using $\lim_{n\to\infty}\left(\frac{2n-j+1}{2n-2j}\right)^n = \exp\left(\frac{j+1}{2}\right)$,

enables (C21) to be recast in a form suitable for calculations that involve large values of $n$ and $j$:

$$eb_0 n\left(\frac{2}{\pi j}\right)^{\frac{1}{4}}\left(\frac{\mathrm{e}}{2}\right)^{\frac{j}{2}}\left(\frac{j}{n}\right)^{\frac{j-3}{2}} \qquad \text{(C22)}$$

These important results demonstrate the rapidity with which the value of the radial part of $p_{if}$ decreases as $n_i - n_f$ increases when $n_i$ is large as the table 1 illustrates.

The result * in table 1 agrees with the value obtained by the more specialised equation (C16).

From equation (9), the decay rate is proportional to $\omega^3 p_{if}{}^2$. Although for, say $j$=$10^{20}$, $\omega^3$ might increase by perhaps $10^{75}$, it is clear from the above table that when $n_i$ is of the order of $10^{30}$, $p_{if}{}^2$ decreases by a much larger factor. The result is that the factor $\omega^3 p_{if}{}^2$ rapidly becomes small as increases significantly making large $\Delta n$ transitions virtually impossible. In fact all states on the first billion or more diagonals are effectively frozen in time and this is more than a sufficient number of states to account for the dark halo of galaxies. It may be that there are sufficient states to make up the required mass using only neutrinos as the eigenstate particles. The rapid decrease in $p_{if}$ is physically understandable in terms of the shapes and behaviour of the Laguerre polynomials for high $n$, which are obtainable through some of their empirical relationships. When $j$ is large, the polynomial contains many oscillations of very similar shape and size, overlapped with the single peaked and relatively wide $j = 1$, $(l = n - 1)$ polynomial. This oscillatory behaviour of the high *j* Laguerre functions leads to almost complete cancellation in the overlap integral. Furthermore for large *j*, the width of the polynomial becomes very large and the amplitude consequently very small because of the normalisation condition. For example for $n = 10^{30} + 10^{20}$, $l = 10^{30}$ (and $j = 10^{20}$) the total spread of the wave function is around 3 x $10^{15}$ m, the wave function amplitude is around 2 x $10^{-28}$, while the oscillation width is about 3 x $10^{-5}$ m. The single peaked eigenstate, $n = 10^{30}$, $l = 10^{30} - 1$, (and $j = 1$) has a width of around 6 x $10^{5}$ m. There are therefore over ten billion oscillations of the $j = 10^{20}$ state under the single peaked $j = 1$ state.

**Acknowledgements**

The author would like to thank Professor N. H. Fletcher of the Institute of Advanced Studies, Australian National University for his helpful comments on this paper.

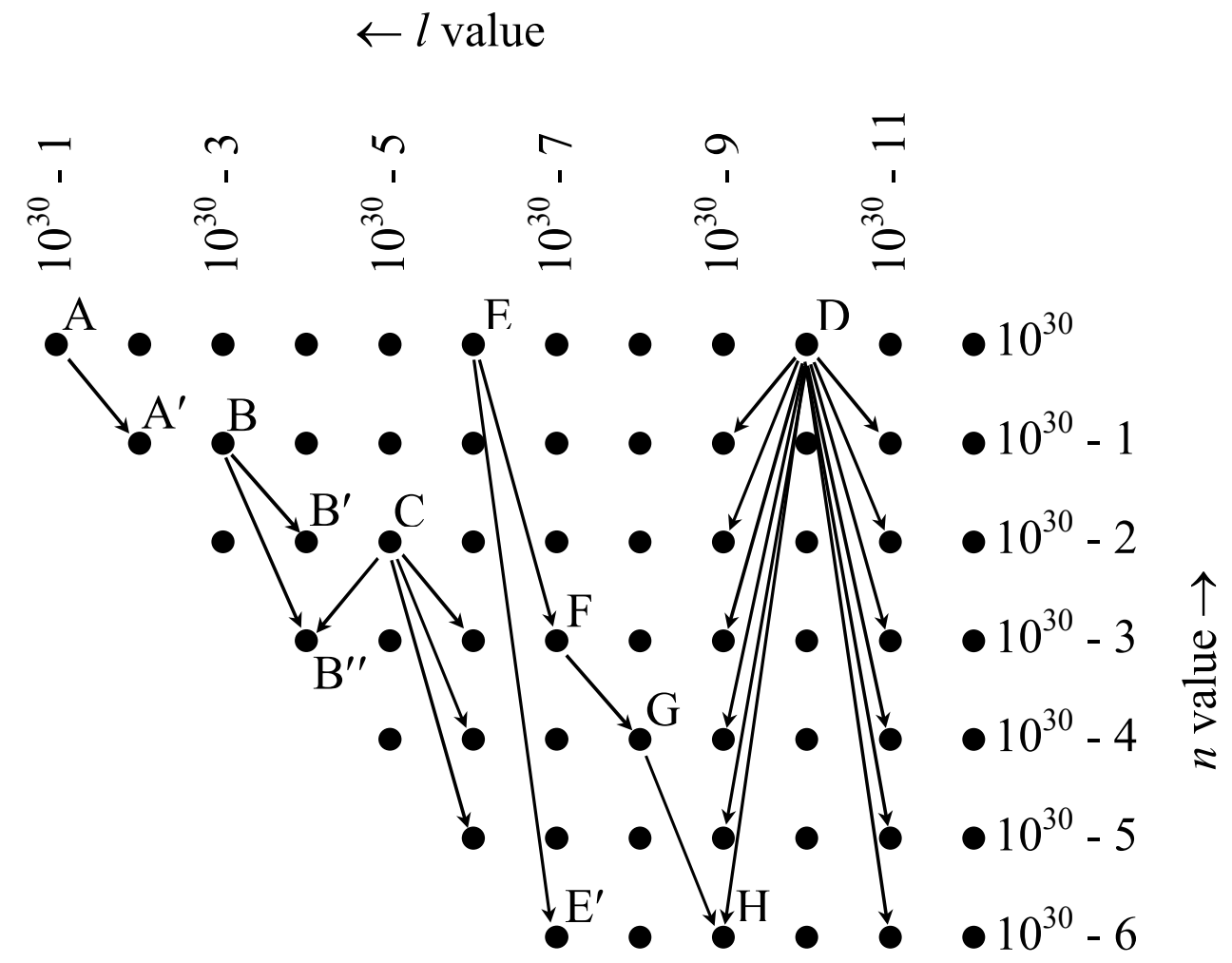

Figure 1. Points representing the high *n* and *l* valued gravitational macro-eigenstates. Each point represents (2 *l* + 1) z - projection sublevels.

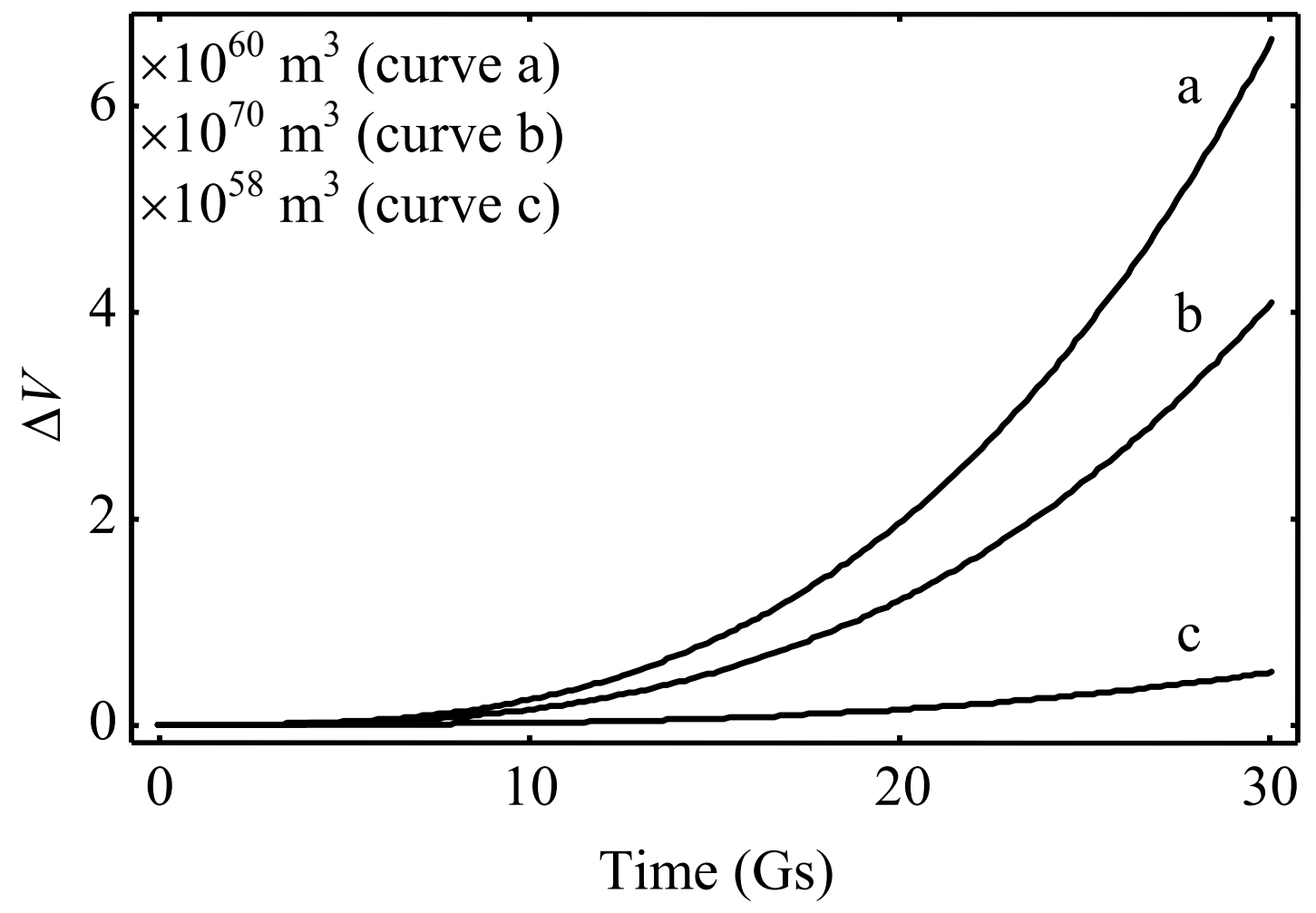

Figure 2. Variation of free particle wave function volume as a function of time and initial volume $\Delta V_i$ for a 3 dimensional Gaussian wave packet; curve a, proton, $\Delta V_i = 125$ fm$^3$; curve b, electron, $\Delta V_i = 125$ fm$^3$; curve c, electron, $\Delta V_i = 10^{-3}$ nm$^3$.

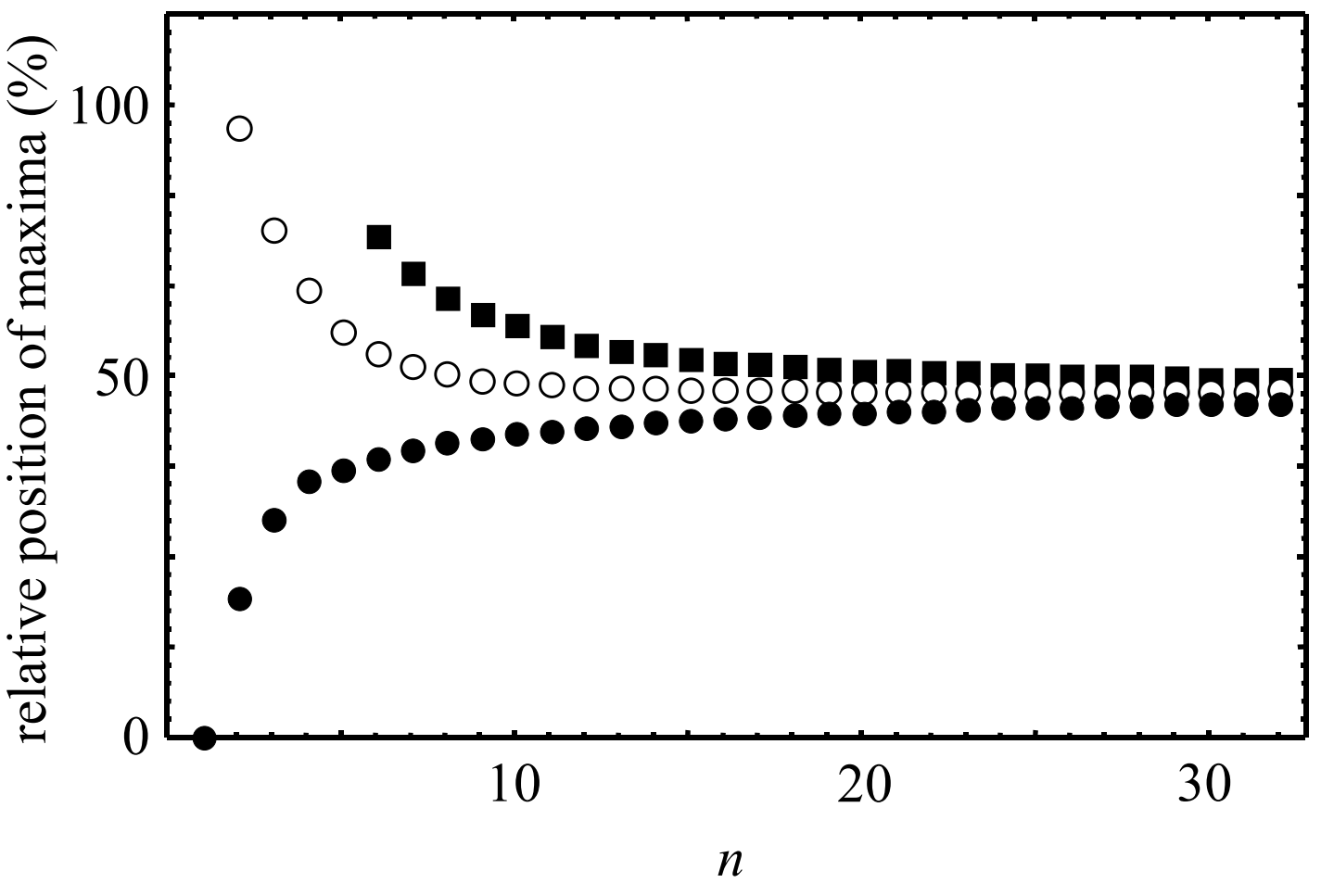

Figure 3. Relative position of the final maximum $r_{\text{finalmax}}$ of the probability function $(R_{nl}r)^2$ compared to the maxima of the truncated functions $(R_{n\ell}r)^2_1$ (●); $(R_{n\ell}r)^2_2$ (■); and $(R_{n\ell}r)^2_3$ (○) as a function of $n$ for $l = 0$.

Table 1 - Radial components of $p_{if}$ for transitions of the type

$$n_i, l_i = n_i - j \text{ to } n_f = n_i - j, l_f = n_f - 1$$

| $n_i$ | $j$ | $\int_0^\infty eR^*_{nf,lf}(r) r^3 R_{ni,li}(r) r^2\, dr$ |
|---|---|---|
| 1000 | 1 | ~$10^6$ $e\, b_0$ |
| 1000 | 5 | ~6 $e\, b_0$ |
| 1000 | 20 | ~3 x$10^{-11}$ $e\, b_0$ |
| 1000 | 100 | ~$10^{-40}$ $e\, b_0$ |
| $10^{30}$ | 1 | ~$10^{60}$ $e\, b_0$ * |
| $10^{30}$ | 5 | ~6 $e\, b_0$ |
| $10^{30}$ | 10 | ~$10^{-71}$ $e\, b_0$ |
| $10^{30}$ | $10^{20}$ | ~$10^{-5\times10^{20}}$ $e\, b_0$ |
| $10^{30}$ | $10^{26}$ | ~$10^{-2\times10^{26}}$ $e\, b_0$ |
| 8 x $10^{33}$ | 5 x $10^{31}$ | ~$10^{-5\times10^{31}}$ $e\, b_0$ |